\documentclass[prd,floats,superscriptaddress,nofootinbib,twocolumn,preprintnumbers]{revtex4}%
\pdfoutput=1  

\usepackage{amssymb}
\usepackage{amsmath}

\usepackage{sans}

\usepackage{graphicx}
\usepackage{graphics}
\usepackage{dcolumn}
\usepackage{color}
\usepackage{rotate}
\usepackage{fancyhdr} 
\usepackage{hyperref}
\usepackage{indentfirst}

\usepackage{umoline}

\def\be{\begin{eqnarray}}
\def\ee{\end{eqnarray}}
\def\ba{\begin{eqnarray}}
\def\ea{\end{eqnarray}}
\def\no{\nonumber}

\def\MBH{M_{\rm BH}}
\def\Msol{M_{\odot}}
\def\muas{\rm\mu arcsec}

\newcommand{\PlotsFolder}{Plots}

\definecolor{darkred}{rgb}{.743,0,0}

\begin{document}
\preprint{IFT-UAM/CSIC-19-72, CERN-TH-2019-081}
\title{Looking for ultralight dark matter near supermassive black holes}

\author{Nitsan Bar}\email{nitsan.bar@weizmann.ac.il}\affiliation{Weizmann Institute of Science,
  Rehovot 7610001, Israel} 

\author{Kfir Blum}\email{kfir.blum@cern.ch}\affiliation{Weizmann Institute of Science,
  Rehovot 7610001, Israel}
\affiliation{Theory department, CERN, CH-1211
  Geneva 23, Switzerland} 
  
  \author{Thomas Lacroix}\email{thomas.lacroix@uam.es}
  \affiliation{Instituto de F\'isica Te\'orica UAM/CSIC, Universidad Aut\'onoma de Madrid, 28049 Madrid, Spain
}
  
\author{Paolo Panci} \email{paolo.panci@unipi.it}
\affiliation{Dipartimento di Fisica, Universit\`{a} di Pisa and INFN, \\ Sezione di Pisa Largo Pontecorvo 3, 56127 Pisa, Italy}
 \affiliation{Theory department, CERN, CH-1211 Geneva 23, Switzerland}
  \affiliation{Laboratori Nazionali del Gran Sasso, Via G.~Acitelli, 22, I-67100 Assergi (AQ), Italy}


\begin{abstract}
Measurements of the dynamical environment of supermassive black holes (SMBHs) are becoming abundant and precise. 
We use such measurements to look for ultralight dark matter (ULDM), which is predicted to form dense cores (``solitons") in the centre of galactic halos. We search for the gravitational imprint of an ULDM soliton on stellar orbits near Sgr A* and by combining stellar velocity measurements with Event Horizon Telescope imaging of M87*. Finding no positive evidence, we set limits on the soliton mass for different values of the ULDM particle mass $m$. 
The constraints we derive exclude the solitons predicted by a naive extrapolation of the soliton-halo relation, found in DM-only numerical simulations, for $2\times10^{-20}~{\rm eV}\lesssim m\lesssim8\times10^{-19}~{\rm eV}$ (from Sgr A*) and $m\lesssim4\times10^{-22}~{\rm eV}$ (from M87*). 
However, we present theoretical arguments suggesting that an extrapolation of the soliton-halo relation may not be adequate: in some regions of the parameter space, the dynamical effect of the SMBH could cause this extrapolation to over-predict the soliton mass by orders of magnitude.
%
\end{abstract}

\maketitle


\section{Introduction}

Supermassive black holes (SMBHs) reside in most galaxies~\cite{Kormendy:1995er,Ferrarese:2004qr,Narayan:2005ie} and their properties (mass, spin, and close environment) are under rapidly improving observational scrutiny. Two SMBHs for which very precise data exists are Sgr A*  
in the dynamical centre of the Milky Way (MW) and M87*  
in the elliptical galaxy M87. 
Near Sgr A*, the orbit of the star S2 in the S star cluster~\cite{Genzel:2003cn,Genzel2010,2017ApJ...837...30G} has been observed along more than a full lap~\cite{2009ApJ...707L.114G,Abuter:2018drb}. 
In the case of M87*, the event horizon telescope (EHT) collaboration has very recently released a breathtaking image of the BH shadow~\cite{Akiyama:2019eap}. 

In this paper we show that measurements of the dynamical environment of SMBHs provide an interesting probe of ultralight dark matter (ULDM)~\cite{Hu:2000ke,Amendola:2005ad,Svrcek:2006yi,Arvanitaki:2009fg,Marsh:2015xka,Lee2018}. 
ULDM gained wide interest partially because in the window $10^{-22}~\text{eV}\lesssim m \lesssim10^{-21}~\text{eV}$, it could alleviate small-scale puzzles facing the dark matter paradigm~\cite{Hu:2000ke,DelPopolo:2016emo,Hui:2016ltb}. This mass range, moreover, defines the absolute lower bound for the possible mass of dark matter. 
At the centre of galactic halos ULDM is expected to develop cored density profiles~\cite{Arbey:2001qi,Lesgourgues:2002hk,Chavanis:2011zi,Chavanis:2011zm,Schive:2014dra,Schive:2014hza,Marsh:2015wka,Calabrese:2016hmp,Chen:2016unw,Schwabe:2016rze,Veltmaat:2016rxo,Hui:2016ltb,Gonzales-Morales:2016mkl,Robles:2012uy,Bernal:2017oih,Mocz:2017wlg,Mukaida:2016hwd,Vicens:2018kdk,Bar:2018acw,Eby:2018ufi,Bar-Or:2018pxz,Marsh:2018zyw,Chavanis:2018pkx,Emami:2018rxq,Levkov:2018kau,Broadhurst:2019fsl,Hayashi:2019ynr,Bar:2019bqz}, commonly referred to as ``solitons",\footnote{A more appropriate term is oscillatons; but we will stick to solitons in what follows.} corresponding to quasi-stationary minimum energy solutions of the equations of motion. The ULDM soliton could be detected given detailed knowledge of the mass distribution in the inner halo. Such detailed view is provided by SMBH precision measurements: in the case of Sgr A*, any additional mass distribution $\delta M$ between the periastron and apoastron of the S2 orbit ($ \sim 0.005$~pc) is constrained at the level of $\delta M/\MBH\lesssim$ few percent \cite{2017ApJ...837...30G}. Measurements of stellar motions at larger distances~\cite{Genzel2010} ($\sim 0.3$~pc) provide more constraints. For M87*, a combination of the EHT measurement with analysis of stellar velocity dispersion at distances of $\lesssim0.5$~kpc can be translated into the constraint $\delta M/\MBH\lesssim10\%$. 
We will show that these observations probe ULDM at a meaningful level.

The problem of a minimally-coupled massive scalar field in the strong gravity regime around a BH (including the superradiance phenomenon~\cite{Brito:2015oca}) was investigated in the literature~\cite{Unruh:1976fm,Detweiler1980,Arvanitaki:2009fg,Arvanitaki:2010sy,Cardoso2011,Ferreira:2017pth,Boskovic:2018rub,Cardoso:2018tly,Hui2019,Benone2019,Chen2019}, recently also in the context of M87*~\cite{Davoudiasl:2019nlo}. Other works~\cite{Hui:2016ltb,Bar:2018acw,Bar-Or:2018pxz} considered the interplay between ULDM and black holes on galactic scales within the Newtonian approximation. Our approach  focuses on the intermediate case, where on the one hand a Newtonian analysis is applicable but on the other hand, the SMBH dominates the dynamics. 

Several other constraints on ULDM have appeared in the literature. The matter power spectrum revealed by Ly-$\alpha$ forest analyses is in tension with $m\lesssim10^{-21}~{\rm eV}$~\cite{Armengaud:2017nkf,Irsic:2017yje,Zhang:2017chj,Kobayashi:2017jcf,Leong:2018opi} (see also~\cite{Bozek:2014uqa,Hlozek:2017zzf}). Rotation curves of low-surface-brightness galaxies (LSBs) also disfavour $m\lesssim 10^{-21}~\text{eV}$~\cite{Bar:2018acw,Bar:2019bqz}, if one accepts the soliton cores predicted by numerical simulations~\cite{Schive:2014dra,Schive:2014hza,Veltmaat:2018dfz}. Independent evidence from rotation curve data against ULDM cores was reported in~\cite{Deng:2018jjz}. Dynamical heating of the MW disk~\cite{Church:2018sro} and a preliminary analysis of stellar streams~\cite{Amorisco:2018dcn} disfavour $m\lesssim10^{-22}~{\rm eV}$. A weaker bound comes from pulsar timing measurements~\cite{Porayko:2018sfa} of scalar metric perturbations induced by ULDM~\cite{Khmelnitsky:2013lxt}, which exclude $m\lesssim10^{-23}~{\rm eV}$. Ref.~\cite{Marsh:2018zyw} showed that a dynamical analysis of a central star cluster in Eridanus-II could potentially probe ULDM up to $m\lesssim 10^{-19}~{\rm eV}$. 

The paper is outlined as follows. 

Section \ref{s:SMBHvsSol} sets the stage for our investigation, introducing a few basic properties of the ULDM soliton, explaining how ballpark numbers for the soliton mass motivate us to look for ULDM near SMBHs, and highlighting a few of the complications we will encounter. 

In section \ref{s:obs} we use observations to search for solitons, considering first Sgr A* (Secs.~\ref{ss:S2} and~\ref{ss:MWcwd}) and then M87* (section \ref{ss:M87EHT}). 
Our goal is to examine how different values of the soliton mass, $M_{\rm sol}$, affect measurements of the SMBH dynamical environment for different assumed values of $m$. 
The observational constraints that we found for Sgr A* and for M87* are compared to theoretical expectations in section \ref{sec:compare} and section \ref{s:M87comp}, respectively. 

Theoretical benchmarks for the soliton are explained in Secs.~\ref{s:howmuch} and~\ref{ss:relax}, with some details postponed to appendix \ref{ss:solbh}. 
The basic benchmark we look at, in section \ref{s:howmuch}, comes from the soliton-host halo relation found in the DM-only numerical simulations of Refs.~\cite{Schive:2014dra,Schive:2014hza}. We find that the soliton-halo relation is tested by the SMBH data in a new range of $m$ compared to previous tests.  
However, the soliton-halo relation involves caveats that prevent us from turning the constraints on $M_{\rm sol}$ into robust exclusion on $m$. First, the solitons we consider must account for the effect of a SMBH, whereas the numerical simulations included only ULDM. Second, the simulations were only run for a limited range of host halo masses and ULDM particle masses, while we explore more massive halos and more massive particles. 

In section \ref{ss:relax} we consider the question of dynamical relaxation. Using the relaxation time estimate of Ref.~\cite{Levkov:2018kau}, combined with observations made in~\cite{Bar:2018acw,Bar:2019bqz}, we show that dynamical relaxation may become a bottleneck for soliton formation for $M_{\rm h}\sim10^{12}~\Msol$ and $m\gtrsim10^{-22}~{\rm eV}$. 
At $m=10^{-19}~{\rm eV}$, for example, the soliton mass prescribed by dynamical relaxation could be an order of magnitude lower than that predicted by naive extrapolation of the scaling relation of~\cite{Schive:2014dra,Schive:2014hza}, even when one ignores the impact of a SMBH.

We summarise our results in section \ref{s:sum}.

We leave some details to appendices. In appendix \ref{ss:solbh} we review the structure of the soliton in the regime where the dynamics is dominated by a SMBH, but where the Newtonian approximation is still valid. A simple approximation for the soliton profile is introduced to facilitate numerical calculations. We calculate the time scale characterising the absorption of a soliton into the SMBH. Our results suggest that over much of the parameter space of interest, Sgr A* and M87* could absorb ULDM too fast to allow for a soliton to be established. 

Finally, in appendix \ref{app:selfint} we outline the parametric region where non-gravitational self-interactions, motivated by axion-like particle models of ULDM, could affect our results.

\section{Setting the Stage}\label{s:SMBHvsSol}
The density profile of a self-gravitating soliton is cored with characteristic radius $x_{\rm c}$ and mass $M_{\rm sol}$ related by (see, e.g.~\cite{Bar:2018acw})\footnote{We define $x_{\rm c}$ as the radius at which the soliton mass density decreases by a factor of 2 compared to its value at the origin~\cite{Schive:2014dra,Schive:2014hza}.} 
\be\label{eq:xMsg} 
\begin{split}
& \mbox{self-gravitating soliton:} \\
& M_{\rm sol}\,x_{\rm c} \approx 2.3\times 10^5\left(\frac{m}{10^{-19}~\rm{eV}}\right)^{-2}~M_\odot\,\rm{pc} \ .
\end{split}
\ee 
A glance at the properties of Sgr A*, with mass $\MBH\approx4\times10^6~\Msol$ dominating the dynamics out to a few pc~\cite{Genzel2010}, suggests that stellar orbit measurements with $\sim10$\% accuracy in the SMBH-dominated region could be sensitive to $m\sim(10^{-20}-10^{-19})~{\rm eV}$ provided that $M_{\rm sol}\sim\MBH$. Interestingly, this ballpark for $M_{\rm sol}$ is consistent with a naive extrapolation (in $m$) of the results of DM-only numerical simulations~\cite{Schive:2014dra,Schive:2014hza}. 

These estimates look promising, and we will see that  measurements of Sgr A*  do lead to constraints that test the extrapolation of~\cite{Schive:2014dra,Schive:2014hza}. However, the presence of the SMBH complicates the situation. 
First of all, when the SMBH dominates the dynamics, the soliton shape is distorted. The characteristic radius becomes independent of $M_{\rm sol}$ and, instead of eq.~(\ref{eq:xMsg}), is given by (see appendix \ref{sss:shape})
\be\label{eq:xMbh} 
\begin{split}
& \mbox{BH-dominated soliton:} \\
& \MBH\,x_{\rm c}\approx 4\times10^{6}\left(\frac{m}{10^{-20}~\rm{eV}}\right)^{-2}~M_\odot\,\rm{pc} \ .
\end{split}
\ee 
The soliton mass is then an independent parameter. Whether the soliton can be probed by stellar orbits, or not, depends on $M_{\rm sol}$ via its relation to the soliton central density, $M_{\rm sol} \approx \pi \rho_0/(G\MBH m^2)^3$. 

The particular complication due to the transition from eq.~(\ref{eq:xMsg}) to eq.~(\ref{eq:xMbh}) does not turn out to be a show stopper, but it does illustrate the impact of the SMBH. We will consider a number of other complications, such as the possible impact of the SMBH on the naive large-$m$ extrapolation (by large-$m$, we mean $m\gtrsim10^{-21}~{\rm eV}$) of the $M_{\rm sol}$ scaling of~\cite{Schive:2014dra,Schive:2014hza}. A key caveat suggested by our findings is that in much of the parameter space where SMBH measurements could naively test ULDM, the soliton may actually be consumed by accretion into the SMBH. Understanding what really happens in this case requires simulating the co-evolving SMBH+ULDM systems, which is beyond the scope of this paper. 
Our take-home message after considering these complications will be that while SMBH measurements open up an interesting avenue to search for ULDM, the theoretical uncertainties are still too large to allow for robust exclusion, at least based on the observables that we analysed.

%

%

%
\section{Looking for ULDM near SMBHs}\label{s:obs}

In this section we derive observational constraints on ULDM solitons from stellar orbits near Sgr A* in the MW (section \ref{sec:MWconst}) and from the EHT measurements and stellar dispersion analyses of M87* (section \ref{ss:M87}). While the results can be read and understood without referring to the technical details of the soliton's structure, the underlying calculations employ tools and results that are explained in subsequent sections. In particular, we use the BH-deformed soliton shape calculation of appendix \ref{sss:shape}, and compare our constraints to theoretical benchmarks which are explained in section \ref{s:howmuch}, section \ref{ss:relax} and appendix \ref{ss:solbh}.

\subsection{Milky Way}\label{sec:MWconst}
We now discuss the constraints obtained from observations around the SMBH in the MW, specifically from the orbit of the star S2 (section \ref{ss:S2}) and from observations of a stellar disk (section \ref{ss:MWcwd}).\footnote{Position and polarization measurements in near infra-red, attributed to flares of Sgr A*~\cite{gravity2018}, may also constrain the properties of the SMBH.} These constrain the $ \sim 0.005 $~pc and $ \sim 0.2 $~pc regions, respectively. Farther away, at the few pc region and outwards, the stellar mass contribution becomes comparable to the BH mass~\cite{2016ApJ...821...44F}. This makes the analysis more involved, beyond the scope of the current work. In section \ref{sec:compare} we compare the constraints from observations to theoretical expectations.

\subsubsection{The orbit of S2}\label{ss:S2}

\begin{figure}[htbp!]
	\centering
	\includegraphics[width=0.45\textwidth]{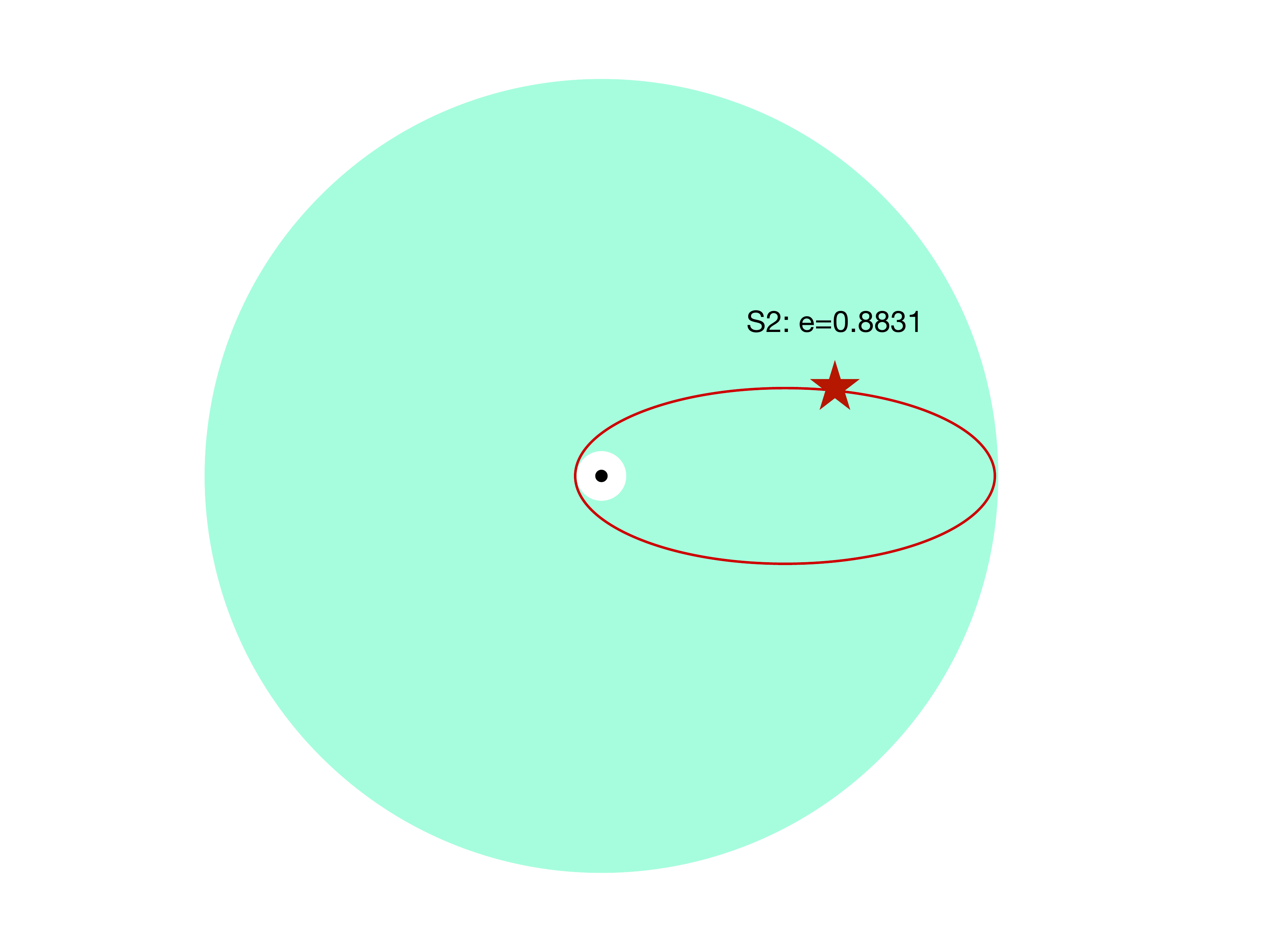}	
	\caption{Schematic view of the S2 elliptic orbit.}\label{fig:S2orbit}
\end{figure}

Precision measurements of the orbits of stars in the S star cluster at the centre of the MW (see, e.g.~\cite{Genzel:2003cn,2017ApJ...837...30G}) are sensitive to the mass distribution near the SMBH. The discriminatory power between an extended mass distribution to an isolated point mass (BH) arises from the eccentricity of the stellar orbit. In figure \ref{fig:S2orbit} we show a schematic view of the elliptic orbit of the B2-type star S2, for which more than a full orbit has been recorded~\cite{2009ApJ...707L.114G}. An extended mass distribution $M^{\rm ext}$, defined as the mass within the green filled shell extending between the periastron and apastron of the orbit, can be constrained independently of the central mass inside of the periastron, shown by the internal white region with a black point representing the SMBH. It should be noted that in the presence of a significant extended mass distribution, the orbit of S2 would exhibit strong precession. In that limit, figure \ref{fig:S2orbit} should be thought of as showing only the osculating orbit of the star.

Our computation follows Ref.~\cite{Lacroix:2018zmg}. There, VLT measurements of the orbit of S2 up to 2016 were used to constrain the distribution of dark mass, which was assumed to exhibit a density spike towards the SMBH. Here, we use the same data to constrain an ULDM soliton. 

\begin{figure}[]
	\centering
	\includegraphics[width=0.45\textwidth]{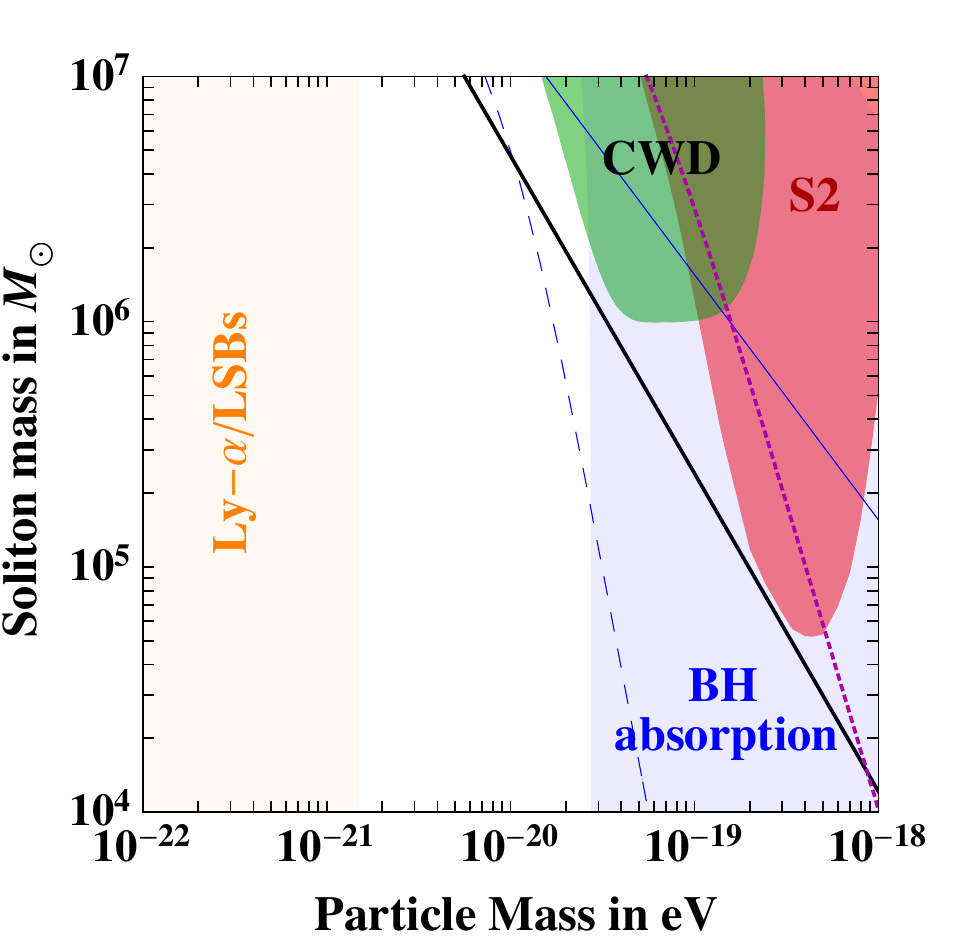}	
	\caption{Constraints on the total mass of an ULDM soliton. The red shaded region, marked  ``S2", is excluded at the 95\,\%~CL by measurements of the star S2 (see section \ref{ss:S2}). The green shaded region, marked ``CWD", is in conflict with observations of a clockwise-rotating disk (see section \ref{ss:MWcwd}). In the blue-shaded region, marked ``BH absorption", the time scale for absorption of the soliton by the SMBH, estimated in section \ref{sss:absorb}, is shorter than $\tau_{\rm U}$. Different lines represent theoretical benchmarks for $M_{\rm sol}$, explained in section \ref{sec:compare}.}\label{fig:S2}
\end{figure}

We use an orbit-fitting procedure as developed in Refs.~\cite{Ghez2008,Gillessen:2008qv,2009ApJ...707L.114G,Boehle2016,2017ApJ...837...30G} and described in detail in Ref.~\cite{Lacroix:2018zmg}. The procedure reconstructs the evolution of the position and velocity of the star on its orbit as a function of time, and constrains the properties of the gravitational potential by fitting the parameters of the model to the data, consistently combined in the likelihood. The data includes right ascension, declination, and radial velocity of S2 from VLT measurements~\cite{2017ApJ...837...30G}.\footnote{Combining the VLT data with the data from the Keck observatory \cite{Boehle2016}, using the procedure of Ref.~\cite{2009ApJ...707L.114G}, only improves the limits on $M_{\mathrm{sol}}$ by an $O(1)$ factor at the price of having 4 additional parameters to reconcile the coordinate systems for both data sets, which leads to additional degeneracies and longer computing times. Therefore, in figure \ref{fig:S2} we only account for the VLT data.} 
The 14 parameters of the problem are the mass of the central object, $M_{\mathrm{BH}}$, and its six phase-space coordinates, namely its distance $R_{0}$, its position on the sky (right ascension $\alpha_{\mathrm{BH}}$, declination $\delta_{\mathrm{BH}}$), and velocity ($v_{\alpha,\mathrm{BH}}$, $v_{\delta,\mathrm{BH}}$, $v_{\mathrm{r,BH}}$), as well as the six phase-space coordinates of the star, and the total soliton mass $M_{\mathrm{sol}}$ which characterizes the mass profile $M^{\rm ext}$ of the soliton. We assume that the SMBH and the soliton are concentric. 

Given an assumed BH mass $\MBH$, ULDM particle mass $m$ and soliton mass $M_{\rm sol}$, we compute $M^{\rm ext}$ using the formulae given in appendix \ref{sss:shape}. For the parameters of the star we consider the initial conditions on the polar radius and angle in the plane of the orbit,  $r_{0}$, $\theta_{0}$, and their corresponding derivatives $\dot{r}_{0}$, $\dot{\theta}_{0}$, as well as the inclination angle $I$ with respect to the plane of the orbit and the standard longitude $\Omega$ of the ascending node.\footnote{Note that we cannot rely on the 6 standard orbital elements that characterize Keplerian orbits since in the presence of an extended mass the orbit is no longer Keplerian. We note however that $I$ and $\Omega$ still define the plane of the orbit even for non-Keplerian motion.}

Following~\cite{Lacroix:2018zmg}, we use \textsf{PyMultiNest}~\cite{Buchner2014}, which relies on the \textsf{MultiNest} multimodal nested Monte Carlo sampling code~\cite{Feroz2009}, to derive the posterior probability distribution of the parameters of the model, in particular $M_{\rm sol}$. We fix the ULDM mass $m$ for a given Monte Carlo run, and scan over $m$ by means of independent runs. 

We find that including an ULDM soliton does not modify the Bayesian evidence in a statistically significant way. Therefore, we derive upper limits on $M_{\rm sol}$ at the 95\% confidence level.\footnote{We use the \textsf{corner.py} Python module \cite{Foreman-Mackey2016}.} The excluded range in the ($m$,$M_{\rm sol}$) plane is shown by the red-shaded region in figure \ref{fig:S2}, marked by ``S2". Total soliton masses down to $\sim 5 \times 10^{4}\, M_{\odot}$ are excluded at the 95\% confidence level for $m \sim 4 \times 10^{-19}\, \rm eV$. It should be noted that above $m \sim 10^{-18}\, \rm eV$, the entire soliton is confined within the pericentre of the orbit of S2, such that the total soliton mass is degenerate with the BH mass. 
%

\subsubsection{A stellar disk}\label{ss:MWcwd}

The orbit reconstruction of S2 and other well-measured members of the S star cluster can probe an ULDM soliton in the inner $r\lesssim0.01$~pc around Sgr A*. This translates into constraints that are particularly strong for $10^{-19}~\text{eV} \lesssim m \lesssim10^{-18}~{\rm eV}$. Additional measurements of stellar kinematics extending to $r\sim0.1-1$~pc provide somewhat less precise estimates of the central mass~\cite{Genzel2010}, but the larger distance probed by these measurements makes them sensitive to smaller values of $m$.

Ref.~\cite{Beloborodov2006} analysed the kinematics of a clockwise-rotating disk (CWD) of stars spanning distances of $\sim0.1-0.3$~pc around Sgr A*. For an 8~kpc distance estimate to Sgr A*, they found $\MBH=(4.3\pm0.5)\times10^6~\Msol$. Combined with the results for the S2 orbit reconstruction, we deduce the constraint:
\be \delta M\left(r<0.3~{\rm pc}\right)&\lesssim&10^6~\Msol \ .\ee
This constraint, translated to the $(m,M_{\rm sol})$ plane, is shown by the green-shaded region, marked by ``CWD" in figure \ref{fig:S2}.

\subsubsection{Comparison of constraints with theoretical expectations}\label{sec:compare}
It is interesting to compare the observational constraints of figure \ref{fig:S2} to theoretical expectations for the soliton mass. As we discuss later on, despite the fact that ULDM solitons were invariably seen to form in numerical simulations, the dominant physical mechanisms controlling soliton formation are not yet understood. This leaves significant room for theoretical uncertainties, the identification and preliminary quantification of which make up some of our results in this paper. To keep the discussion concise, we postpone most of the details to sections \ref{s:howmuch}, \ref{ss:relax} and \ref{sss:absorb}. Here we highlight the main results, as follows:
\begin{enumerate}
	\item The thin solid blue line in figure \ref{fig:S2} shows the value of $M_{\rm sol}$ predicted by eq.~(\ref{eq:solhost}) using $M_{\rm h}=1.54\times10^{12}~\Msol$ for the MW halo~\cite{10.1093/mnras/stw2759,Watkins2019}. This benchmark ignores possible effects due to the SMBH. 
	\item The thin dashed blue line shows a more conservative prediction, obtained from eq.~(\ref{eq:solhalK}) subject to the assumption that the SMBH formation preceded the soliton formation: the reasoning behind this prediction is summarised in section \ref{s:howmuch}. 
Note that if, on the other hand, the soliton formed early preceding the SMBH, then the SMBH formation may actually attract more ULDM mass into the soliton; in which case $M_{\rm sol}$ could exceed not only the thin dashed line, but also the solid blue line. 
	\item The thick solid black line shows the constraint on the soliton mass, that arises if one assumes that the soliton is dominantly formed in the kinetic regime via dynamical relaxation. This line is computed using eq.~(\ref{eq:k2ms2}) following the reasoning presented in section \ref{ss:relax}. It is equivalent to the central value of the blue band in figure \ref{fig:solrelax}. The SMBH is ignored in this computation.
	\item In the blue-shaded region, occupying approximately half of the plot\footnote{With the exception of the upper-right corner, where due to the degeneracy between $M_{\rm sol}$ and $\MBH$, the latter is compatible with zero.} at $m>3\times10^{-20}~{\rm eV}$, a rough estimate of the time scale for absorption of the soliton by the SMBH, computed in section \ref{sss:absorb}, is shorter than $\tau_{\rm U} \simeq 13.8$~Gyr. In this region, the soliton may be entirely eaten by the BH and precise determination of the dynamics would require simulating the co-evolving SMBH and ULDM systems.
	\item Finally, in the region above the thick-dotted magenta line, axion-like particle models of ULDM, where initial field misalignment in a cosine potential determines the ULDM relic abundance, predict that non-gravitational self-interactions could modify the soliton solution. Note that the SMBH causes the field to compress, making non-linearities more important than in the self-gravitating soliton case. We also note, however, that the precise range in $m$ where self-interactions become important depends strongly on the initial misalignment: a $\sim10$\% tuning in the initial conditions would shift the dashed magenta line up by a factor of $\sim100$. The details are given in appendix \ref{app:selfint}. 
\end{enumerate}

\subsection{Messier 87}\label{ss:M87}

In this section we derive observational constraints on ULDM solitons by comparing the EHT measurement and stellar kinematics measurements of M87*, the SMBH in M87. In section \ref{ss:M87EHT} we present the constraint and in section \ref{s:M87comp} we discuss its implications.

\subsubsection{EHT BH shadow vs. stellar kinematics in M87}\label{ss:M87EHT}

The EHT collaboration has recently reported an image of the shadow of the SMBH M87*~\cite{Akiyama:2019eap}.\footnote{For a previous discussion see~\cite{Lu:2014zja} and references therein.} 
The BH shadow observed by the EHT  
translates into a gravitational angular radius~\cite{Luminet:1979nyg,Chandrasekhar:1985kt,Falcke:1999pj,Takahashi:2004xh}  
\be \theta_{\rm g}&=&\frac{G\MBH}{c^2D}
=\left(3.8\pm0.4\right)~\muas \ ,\label{eq:eht}\ee
where $D$ is the distance to the BH. 
In what follows we will assume a fiducial distance of $D=17$~Mpc, in which case one finds $\MBH=\left(6.6\pm0.7\right)\times10^9~\Msol$. 

The EHT result allows for a new test of the mass distribution in the inner region of the galaxy.\footnote{Here we consider the constraints on the mass distribution far away from the BH horizon, in the weak field regime. The mass distribution in the strong field regime could, in principle, be tested too~\cite{Lacroix:2012nz}.}
Ref.~\cite{Gebhardt:2011yw} analysed stellar kinematics, where the most detailed and precise data used in the analysis fell in the range $\theta_*=(2.5''-11'')$. Combined with the value of $D$ used by~\cite{Gebhardt:2011yw}, their result for $\MBH$ translates into $GM(\theta_*)/(c^2D)=\left(3.6\pm0.2\right)~\muas$. 
Comparing eq.~(\ref{eq:eht}) to the results of~\cite{Gebhardt:2011yw}, we see that an additional mass distribution within $\theta_*>\theta_{\rm g}$, parametrised by $\delta M(\theta_*)$, is constrained by: 
\be\frac{\delta M(\theta_*)}{\MBH}&=&\frac{GM(\theta_*)}{c^2D}\frac1{\theta_{\rm g}}-1=-0.04\pm0.11 \ .\label{eq:dM87}\ee
For simplicity, we interpret eq.~(\ref{eq:dM87}) to hold for $\theta_*=(2.5''-11'')$, understanding that additional information could be deduced in a more detailed analysis extending to somewhat larger or smaller $\theta_*$. 

\begin{figure}[htbp!]
	\centering
	\includegraphics[width=0.45\textwidth]{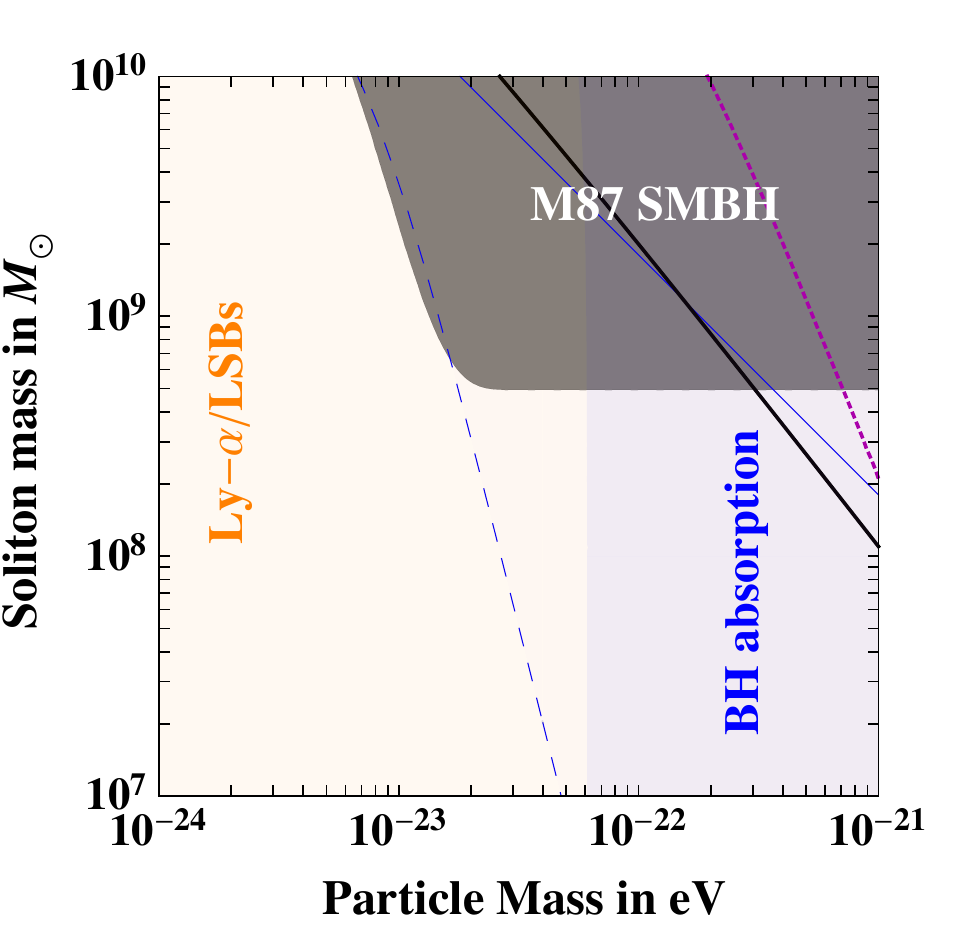}	
	\caption{Constraints on an ULDM soliton derived from combining EHT BH shadow measurements~\cite{Akiyama:2019eap} with the stellar dispersion analysis of Ref.~\cite{Gebhardt:2011yw}. The grey shaded region, marked ``M87 SMBH", is excluded, as there the soliton would spoil the agreement between~\cite{Akiyama:2019eap} and~\cite{Gebhardt:2011yw} on the value of $\MBH$. Various theoretical benchmarks are shown and explained in section \ref{s:M87comp}.}\label{fig:M87}
\end{figure}

Given the values of $m$ and $\MBH$, we can calculate the mass profile of an ULDM soliton of total mass $M_{\rm sol}$ and use eq.~(\ref{eq:dM87}) to derive constraints in the $(m,M_{\rm sol})$ plane. The result of this exercise is shown in figure \ref{fig:M87}. In the grey-shaded region, marked ``M87 SMBH", the soliton mass contribution exceeds the upper limit defined by eq.~(\ref{eq:dM87}) for $\theta_*=11''$. To compute the plot we use the fiducial values $\MBH=6.6\times10^9~\Msol$ and $D=17$~Mpc; varying these fiducial values within the range $D=(17\pm2)~{\rm Mpc}$ and $\MBH=\left(6.6\pm1\right)\times10^9~\Msol$ does not affect the results appreciably.   

The uncertainty in eq.~(\ref{eq:dM87}) is dominantly systematic, and there is room to regard it with care. 
We comment that Ref.~\cite{Walsh:2013uua} considered gas-dynamical models at smaller radii compared to those entering the stellar dispersion of~\cite{Gebhardt:2011yw}, and found the result $GM(\theta_*)/(c^2D)=\left(1.9^{+0.5}_{-0.4}\right)~\muas$: a factor of two lower than that found by~\cite{Gebhardt:2011yw}. 
The discrepancy between the gas models of Ref.~\cite{Walsh:2013uua} and the stellar kinematics analysis of Ref.~\cite{Gebhardt:2011yw} could be due to uncertainties in modelling the gas distribution, e.g. the inclination of the gas disc. 
Prior to the EHT measurement, one might have argued that the factor of two mismatch between the results of Refs.~\cite{Gebhardt:2011yw} and \cite{Walsh:2013uua} could in principle come from a dark halo contribution. The EHT closes this window of opportunity with a result at $\theta_{\rm g}\ll\theta_*$ that confirms the conclusions of Ref.~\cite{Gebhardt:2011yw}. In our analysis we therefore used the stellar dispersion analysis of Ref.~\cite{Gebhardt:2011yw}. To get a quick estimate of the impact of changing the allowed soliton mass from $\delta M/\MBH<0.07$, shown in eq.~(\ref{eq:dM87}), to $\delta M/\MBH<X$, one can multiply the horizontal lower boundary of the grey-shaded area (equal to $M_{\rm sol}\approx5\times10^8~\Msol$ in eq.~(\ref{eq:dM87})) by a factor of $X/0.07$.

\subsubsection{Comparison of constraints with theoretical expectations}\label{s:M87comp}
Here we briefly compare the observational constraint with theoretical expectations.

The thin solid blue line in figure \ref{fig:M87} shows the value of $M_{\rm sol}$ predicted by eq.~(\ref{eq:solhost}) using $M_{\rm h}=2.4\times10^{12}~\Msol$ for M87~\cite{Wu:2005wi}. 
The thin dashed blue line shows the more conservative prediction that is obtained from eq.~(\ref{eq:solhalK}), as explained in section \ref{s:howmuch}.  
In the region above the thick-dotted magenta line, non-gravitational self-interactions may be important, as discussed in appendix \ref{app:selfint}.
In the blue-shaded region, at $m\gtrsim6\times10^{-23}~{\rm eV}$, the time scale for absorption of the soliton by the SMBH, estimated in appendix \ref{sss:absorb}, is shorter than $\tau_{\rm U}$. The solid black line shows the constraint on the soliton mass, which arises if one assumes that the soliton is dominantly formed in the kinetic regime via dynamical relaxation (see section \ref{ss:relax}). 

\section{How much mass in the soliton?}\label{s:howmuch}

The DM-only numerical simulations of Ref.~\cite{Schive:2014dra,Schive:2014hza} related the soliton mass, $M_{\rm sol}$, to the mass of the galactic halo in which it occurs, $M_{\rm h}$:
\be\label{eq:solhost} M_{\rm sol}\approx6.5\times10^8\left(\frac{m}{10^{-22}~\rm eV}\right)^{-1}\left(\frac{M_{\rm h}}{10^{11}~\Msol}\right)^{\frac{1}{3}}M_{\odot} \ .\ee
Ref.~\cite{Bar:2018acw,Bar:2019bqz} showed that for DM-only halos, eq.~(\ref{eq:solhost}) is equivalent to the statement:
\be\label{eq:solhalK}
\frac{K}{M}\Big|_{\rm soliton}&\approx&\frac{K}{M}\Big|_{\rm halo} \ .
\ee
Namely, the numerical simulations of~\cite{Schive:2014dra,Schive:2014hza} are finding that the kinetic energy per particle is the same for ULDM particles in the large-scale host halo and in the central soliton. 

Independent simulations by Ref.~\cite{Veltmaat:2018dfz} showed a result consistent with eq.~(\ref{eq:solhost}). The simulations of Ref.~\cite{Mocz:2017wlg} found a different scaling, but as discussed in~\cite{Bar:2018acw} it remains to be seen if the initial conditions employed in that simulation biased the result. To our knowledge, a direct comparison between the soliton growth found in the simulations of Ref.~\cite{Levkov:2018kau} and those of~\cite{Schive:2014dra,Schive:2014hza} had not yet been made. There is, therefore, room for caution in accepting the soliton-halo relation. We expect that the theoretical situation will become clearer in the near future as different groups test the validity of eq.~(\ref{eq:solhost}). 

We have used eq.~(\ref{eq:solhost}) as an illustrative benchmark with which observational constraints in the $(m,M_{\rm sol})$ plane can be compared. 
However, it is important to note that the reference to Eqs.~(\ref{eq:solhost}-\ref{eq:solhalK}) entails two significant caveats:
\begin{enumerate}
\item Eqs.~(\ref{eq:solhost}-\ref{eq:solhalK}) were found in DM-only simulations while here we consider galaxies which host a SMBH.
\item Eqs.~(\ref{eq:solhost}-\ref{eq:solhalK}) express an empirical result that has only been tested  in numerical simulations of halos in the mass range $M_{\rm h}\sim(10^9-5\times 10^{11})~\Msol$ and for ULDM particle mass $m\approx10^{-22}~{\rm eV}$. In contrast, here we consider more massive halos, $M_{\rm h}\gtrsim10^{12}~\Msol$, and more massive particles, $m\geq10^{-22}~{\rm eV}$. For such $M_{\rm h}$ and $m$, using Eqs.~(\ref{eq:solhost}-\ref{eq:solhalK}) involves significant extrapolation.
\end{enumerate}

In the rest of this section we discuss the effect of the SMBH on ULDM energetics in the soliton region, with possible implications on the soliton formation. Some technical details are postponed to appendix \ref{ss:solbh}. The question of dynamical relaxation is considered in section \ref{ss:relax}.\\

While eq.~(\ref{eq:solhalK})---just like eq.~(\ref{eq:solhost})---was not tested in simulations that include an external baryonic contribution to the gravitational potential, it is suggestive to consider it as evidence for kinetic equilibration between the ULDM reservoir in the halo and in the soliton. This kinetic equilibration is unlikely to represent true steady-state equilibrium, but it could correspond to a bottleneck in the soliton formation which slows down once the soliton grows to saturate eq.~(\ref{eq:solhalK}). Assuming that this is the case, we could use eq.~(\ref{eq:solhalK}) to estimate the outcome of dynamical heating of the inner region of the halo due to the SMBH,\footnote{Ref.~\cite{Bar:2019bqz} considered the related effect of stellar and gas mass components in low surface-brightness galaxies.} which would affect the value of $K/M|_{\rm soliton}$ on the LHS of eq.~(\ref{eq:solhalK}).

In the limit of SMBH dominance, the soliton specific kinetic energy becomes $K/M|_{\rm soliton}\to A^2/2$ where $A=G\MBH m$ (see appendix \ref{sss:shape}, in particular eq.~(\ref{eq:A})). Therefore, if $A^2/2>K/M|_{\rm halo}$, then eq.~(\ref{eq:solhalK}) cannot be satisfied. This would imply that the soliton formation is halted by the dynamical heating due to the SMBH. 

In Figs.~\ref{fig:S2} and~\ref{fig:M87} the soliton mass predicted by eq.~(\ref{eq:solhalK}), including the effect of the SMBH on the LHS, was shown by the thin dashed blue lines.

The discussion above is relevant if the SMBH preceded the formation of the soliton. In reality, the timing could be reversed: the soliton may form early and precede the SMBH (and even, perhaps, help to seed the SMBH). If this latter ordering is the relevant one, then we see no reason to expect that SMBH formation would quench the soliton (that is, apart from the possibility that the SMBH could eat-up the soliton altogether: this is considered in appendix \ref{sss:absorb}). Adiabatic contraction could actually increase the soliton mass above the prediction of eq.~(\ref{eq:solhost}). An example illustrating related dynamics was considered in Ref.~\cite{2017arXiv171201947C}, which simulated the scenario where a population of massive point particles (``stars") was added to a halo containing an existing soliton, and the system was then allowed to evolve dynamically. The stars in the simulation flowed to the centre of the halo, causing the soliton to absorb additional mass from the large-scale ULDM host halo. After a Hubble time, the system attained near steady-state containing a soliton that was {\it more massive} than the prediction of eq.~(\ref{eq:solhost}).

\section{The question of the relaxation time}\label{ss:relax} 
In this section we consider the question of dynamical relaxation, which, for massive halos with $M_{\rm h}\sim10^{12}~\Msol$, may imply a bottleneck for the soliton mass for $m\gtrsim10^{-21}~{\rm eV}$. Throughout this section we ignore the dynamical impact of a SMBH on the soliton: tackling the combined problem of soliton formation via dynamical relaxation alongside a simultaneously-forming SMBH is beyond the scope of this paper. Our analysis could therefore be justified if the soliton forms before the SMBH. Beyond the particular focus of the current work, our analysis in this section may be more generally useful for the understanding of DM-only numerical simulations.

Ref.~\cite{Hui:2016ltb} pointed out that interference patterns in ULDM facilitate dynamical relaxation by acting as quasi-particles of mass $\tilde m\sim\rho/(m\sigma)^3$, where $\rho$ is the mass density and $\sigma$ is the characteristic velocity dispersion (the precise matching to quasi-particles was derived in Ref.~\cite{Bar-Or:2018pxz}). The time scale for dynamical relaxation due to two-body gravitational interactions between the quasi-particles is $\tau\propto\tilde m/(\rho l^2\sigma\Lambda)\propto m^3\sigma^6/(G^2\rho^2\Lambda)$, where $l=G\tilde m/\sigma^2$ is the impact parameter for gravitational collisions and $\Lambda$ is the Coulomb logarithm.
This scaling was verified in Ref.~\cite{Levkov:2018kau} using numerical simulations, which found that the relaxation time in a region of spatial extent $R\gg1/(m\sigma)$ is given by
\be\label{eq:trel}\tau&=&b\frac{\sqrt{2}}{12\pi^3}\frac{m^3\sigma^6}{G^2\rho^2\ln\left(m\sigma R\right)} \ ,\ee
with $b\approx0.7$. Within time $\tau$, a soliton forms in an ensemble of ULDM particles which initially contained no soliton. 

Consider an initial NFW halo density profile~\cite{Navarro:1996gj} characterised by the radius parameter $R_{\rm s}$, a concentration parameter $c$ and a halo virial mass $M_{\rm h}$ (defined as the mass within $R\leq cR_{\rm s}$),
\be\label{eq:nfw}\rho_{\rm NFW} (r) &=&\frac{M_{\rm h}}{4\pi\bar c}\frac{1}{r\left(r+R_{\rm s}\right)^2} \ ,\ee
with $\bar c=\ln(1+c)-c/(1+c)$. If we set $\sigma^2=GM(R)/R$, we can calculate $\tau$ as a function of the mass coordinate $M(R)=4\pi\int_0^Rdr \, r^2\rho_{\rm NFW}(r)$. Inverting this relation gives the mass $M_{\rm relax}$ contained in the relaxed region as a function of the time available for relaxation.

\begin{figure}[htbp!]
	\centering
	\includegraphics[width=0.45\textwidth]{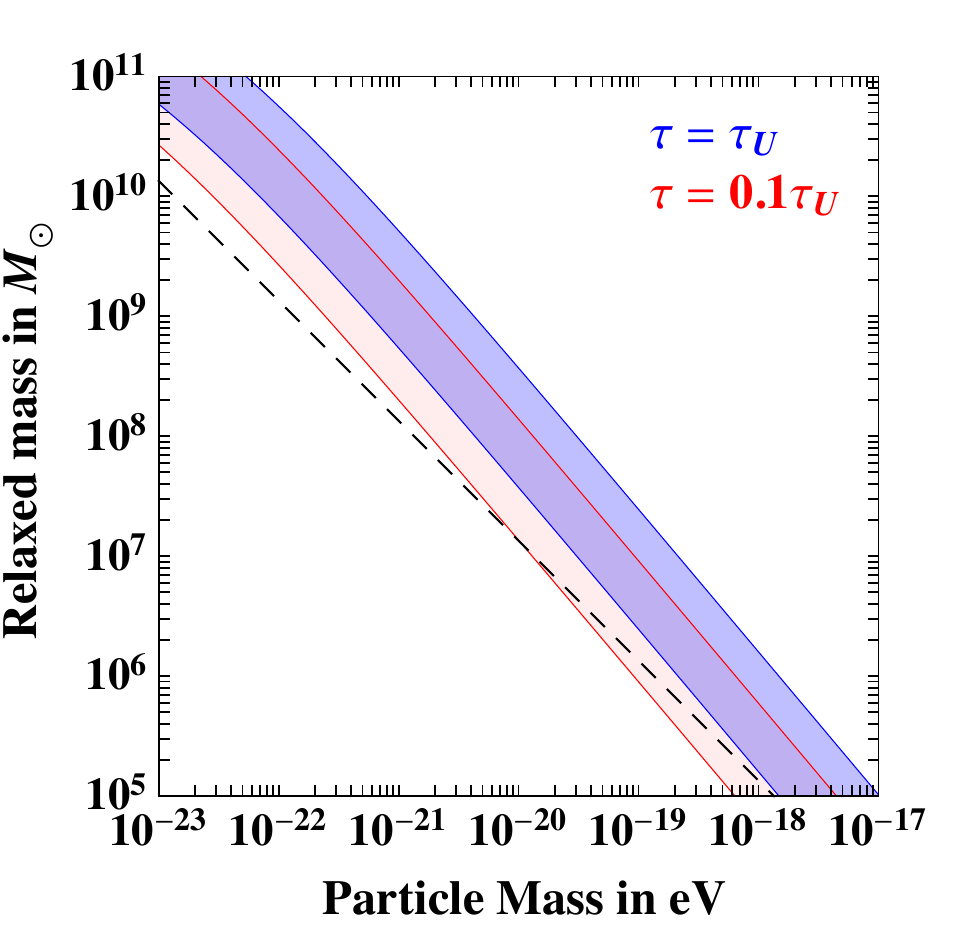}
	\caption{The mass contained in the dynamically relaxed region of an $M_{\rm h}=10^{12}~\Msol$ NFW halo, given $\tau_{\rm U}$ (blue) or $0.1~\tau_{\rm U}$ (red) to relax. The width of each shaded band shows $M_{\rm relax}$ for different values of the halo concentration parameter $c$ and scale parameter $R_{\rm s}$, varying in the range $c=(5,30)$ and $R_{\rm s}=(5,20)$~kpc. The dashed black line shows the soliton mass expected from numerical simulations by extrapolating eq.~(\ref{eq:solhost}) to $m>10^{-22}~{\rm eV}$.}\label{fig:relax}
\end{figure}

In figure \ref{fig:relax} we show the mass contained in the relaxed region of an $M_{\rm h}=10^{12}~\Msol$ NFW halo, given $\tau_{\rm U}$ (blue) or $0.1~\tau_{\rm U}$ (red) 
to relax. The width of each shaded band shows $M_{\rm relax}$ for different values of the halo concentration parameter $c$ and scale parameter $R_{\rm s}$, varying in the range $c=(5,30)$ and $R_{\rm s}=(5,20)$~kpc~\cite{BoylanKolchin:2009an,Piffl:2014mfa}. For comparison, the dashed black line shows the soliton mass expected from extrapolating  the numerical simulations of~\cite{Schive:2014dra,Schive:2014hza,Veltmaat:2018dfz} to $m>10^{-22}~{\rm eV}$. 

\begin{figure}[htbp!]
	\centering
	\includegraphics[width=0.45\textwidth]{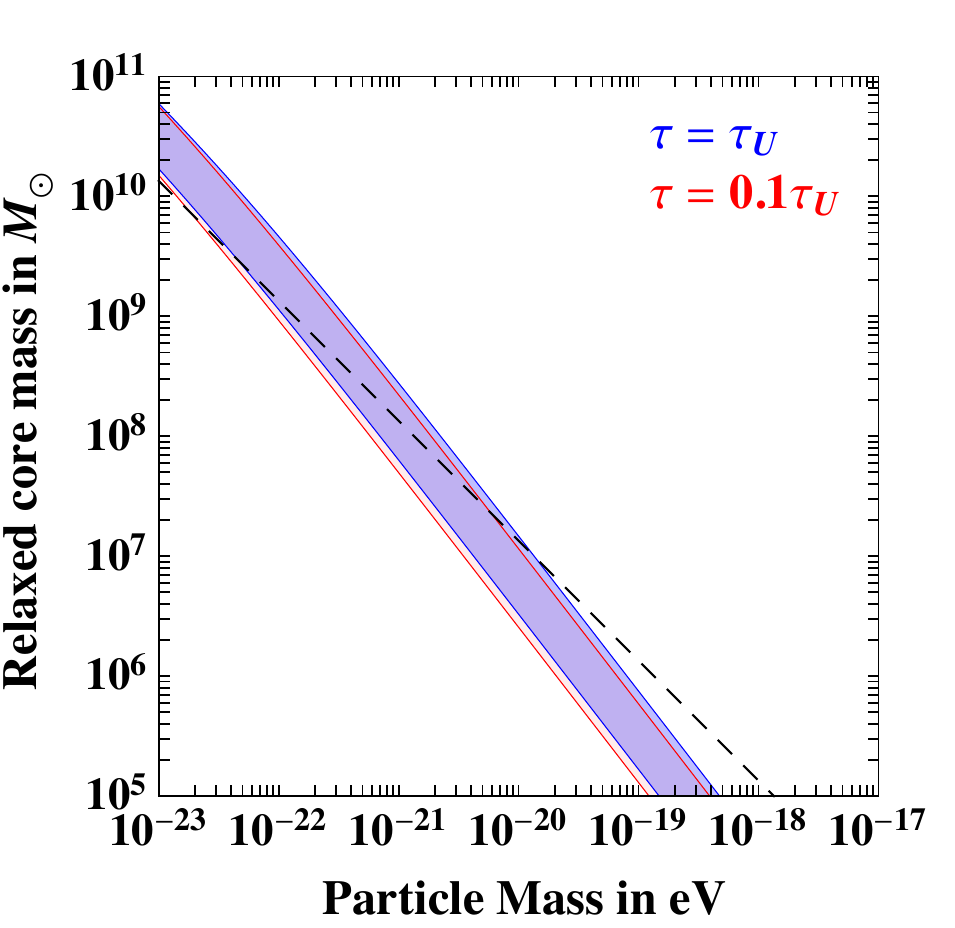}
	\caption{Soliton mass upper bound derived from eq.~(\ref{eq:k2ms2}), where $\sigma^2$ is the velocity dispersion in the dynamically relaxed mass of an $M_{\rm h}=10^{12}~\Msol$ NFW halo, given $\tau_{\rm U}$ (blue) or $0.1~\tau_{\rm U}$ (red) to relax. The width of the shaded bands comes from varying the halo morphology as in figure \ref{fig:relax}.}\label{fig:solrelax}
\end{figure}

The mass $M_{\rm relax}$ could be considered as an upper bound on the mass of the soliton forming in a halo in the kinetic regime~\cite{Hui:2016ltb}. However, $M_{\rm sol}<M_{\rm relax}$ is probably a weak upper bound. If we use eq.~(\ref{eq:solhalK}) to estimate the soliton by equating
\be\label{eq:k2ms2} \frac{K}{M}\Big|_{\rm soliton}=\frac{\sigma^2}{2} \ ,\ee
where $\sigma^2$ is the velocity dispersion at $M_{\rm relax}$, we obtain a stronger (and potentially more realistic) upper bound. This comes from noting that, for a self-gravitating soliton, eq.~(\ref{eq:k2ms2}) translates into $M_{\rm sol}\approx (4.3/Gm)\sqrt{(K/M)|_{\rm sol}}=  (4.3/Gm)\sigma/\sqrt{2}$~\cite{Bar:2018acw}. This upper bound is shown in figure \ref{fig:solrelax}.

It is interesting to consider the parametric scaling of the soliton mass upper bound derived from eq.~(\ref{eq:k2ms2}). 
For massive halos like MW and M87, with $M_{\rm h}\sim10^{12}~\Msol$, and for $m>10^{-21}~{\rm eV}$ or so, the mass contained in the relaxed region within a Hubble time is much smaller than $M(R_{\rm s})$. The density, enclosed mass and velocity dispersion in the relevant region can then be approximated by $\rho_{\rm NFW}\approx M_{\rm h}/(4\pi\bar c R_{\rm s}^2R)$, $M\approx M_{\rm h}R^2/(2\bar c R_{\rm s}^2)$ and $\sigma^2\approx GM_{\rm h}R/(2\bar c R_{\rm s}^2)$. %
Inserting this into eq.~(\ref{eq:trel}) we have:\footnote{We chose the presentation of eq.~(\ref{eq:tauapp}) to highlight the fact that the halo parameters enter only via the combination $M_h/\left(\bar cR_s^2\right)$.}
\be\label{eq:tauapp}\tau&\approx&\frac{8}{3\pi}\frac{GM_{\rm h}}{\bar cR_{\rm s}^2}\frac{\left(\frac{\bar cR_{\rm s}^2M}{M_{\rm h}}\right)^{\frac{5}{2}}m^3}{\ln\left[\frac{\sqrt{2}GM_{\rm h}}{\bar cR_{\rm s}^2}\left(\frac{\bar cR_{\rm s}^2M}{M_{\rm h}}\right)^{\frac{3}{2}}m^2\right]} \ .\ee
In the regime of validity of eq.~(\ref{eq:tauapp}), the mass contained in the relaxed region at fixed $\tau$ scales as $M_{\rm relax}\propto \tau^{2/5}\left(M_{\rm h}/(\bar cR_{\rm s}^2)\right)^{{3}/{5}}m^{-{6}/{5}}$, up to a logarithmic correction. 
Eq.~(\ref{eq:k2ms2}) then leads to $M_{\rm sol}\propto \tau^{{1}/{10}}({M_{\rm h}}/{(\bar cR_{\rm s}^2)})^{{2}/{5}}m^{-{13}/{10}}$, up to a logarithmic correction. Noting that $R_{\rm s}\propto M_{\rm h}^{{1}/{3}}$, and ignoring the dependence on the concentration parameter $c$, we have $M_{\rm sol}\propto \tau^{{1}/{10}}M_{\rm h}^{{2}/{15}}m^{-{13}/{10}}$. The $\sim m^{-{13}/{10}}$ scaling at $m>10^{-21}~{\rm eV}$ is apparent in figure \ref{fig:solrelax} when comparing the shaded band (coming from eq.~(\ref{eq:k2ms2})) to the dashed line (coming from the naive extrapolation of eq.~(\ref{eq:solhost})) which scales as $\sim m^{-1}$.
%

For $m<10^{-21}~{\rm eV}$, the relaxed region extends up to $R\sim R_{\rm s}$ and the small $R$ expansion of eq.~(\ref{eq:tauapp}) does not apply. Instead, $\sigma^2$ becomes approximately independent of $R$ ($\mathrm{d}\sigma^2/\mathrm{d}R$ vanishes at $R\approx2.16R_{\rm s}$). Because $\sigma$ is approximately independent of $R$, it follows that the soliton determined by eq.~(\ref{eq:k2ms2}) is approximately independent of the time available for relaxation. In this regime the upper bound prescribed by eq.~(\ref{eq:k2ms2}) approximately coincides with eq.~(\ref{eq:solhalK}). 

Finally, note that eq.~(\ref{eq:trel}) and the corresponding dynamical relaxation bottleneck apply to the kinetic regime $R\gg1/(m\sigma)$. These considerations may significantly over-estimate the actual time scale for soliton formation, if the initial conditions admit small velocity dispersion, as would be the case in cosmological halos that decouple from the Hubble flow before virialisation~\cite{Levkov:2018kau}.

\section{Summary}\label{s:sum}

Measurements of the mass and dynamical environment of  SMBH are becoming increasingly precise. Two SMBHs where precision measurements have become available are Sgr A* ($\MBH\sim4\times10^6~\Msol$, via stellar orbits) and M87* ($\MBH\sim6\times10^9~\Msol$, via BH shadow imaging and stellar velocity dispersion). The SMBH measurements provide clean probes of the mass distribution in the region where the BH dominates the dynamics. 

We study the implications of the SMBH measurements for ULDM. Analytical arguments and numerical simulations predict that ULDM should form dense cores (``solitons") in the centre of galactic halos. 
We present a search for the gravitational imprint of an ULDM soliton with mass $M_{\rm sol}$ on the orbit of the star S2. We also consider constraints from the observations of a stellar disk. We find no evidence for $M_{\rm sol}>0$ and use the data to derive constraints in the ($m,M_{\rm sol}$) plain, where $m$ is the ULDM particle mass (see figure \ref{fig:S2}). 
We then use stellar velocity dispersion analyses combined with the Event Horizon Telescope (EHT) measurement of M87* to constrain the presence of a soliton, which could manifest itself as excess mass in the velocity dispersion data as compared to the EHT determination of $\MBH$ (see figure \ref{fig:M87}).

DM-only numerical simulations suggest a scaling relation~\cite{Schive:2014dra,Schive:2014hza} that predicts $M_{\rm sol}$ given the host halo mass $M_{\rm h}$ and the particle mass $m$. The observational constraint we find from SgrA*, combining the S2 (Sec.~\ref{ss:S2}) and CWD (Sec.~\ref{ss:MWcwd}) measurements, exclude a naive extrapolation of this relation for $2\times10^{-20}~{\rm eV}\lesssim m\lesssim8\times10^{-19}~{\rm eV}$ (see figure \ref{fig:S2}). Similarly, from M87*, combining EHT and stellar kinematics measurements (Sec.~\ref{ss:M87EHT}), the range $ m\lesssim4\times10^{-22}~{\rm eV}$ is naively excluded (figure \ref{fig:M87}). 

However, a number of theoretical arguments lead us to expect that the naive extrapolation of the soliton-halo relation is not adequate. 
The most significant caveats are the process of soliton absorption by the SMBH, and the question of dynamical relaxation. Both caveats become more pronounced as $m$ is increased, and suggest that a naive extrapolation of the scaling relation of~\cite{Schive:2014dra,Schive:2014hza} to $m\gtrsim10^{-21}~{\rm eV}$ could over-predict the soliton mass by orders of magnitude. Thus, while the SMBH analysis is a potentially interesting discovery tool for ULDM, the theoretical uncertainties are too large to make it a robust exclusion tool. \\

\textit{Note added}: while this manuscript was prepared for publication, Ref.~\cite{Desjacques2019} appeared, discussing stellar dynamics near SMBHs in the presence of ULDM and claiming the strong constraint $m> 10^{-18}~{\rm eV}$. This conclusion appears to assume that the soliton-host halo relation of~\cite{Schive:2014dra,Schive:2014hza} remains valid up to $m\sim 10^{-18}~{\rm eV}$. However, we note that as explained here, the naive extrapolation of the soliton-host halo relation up to $m\gtrsim10^{-21}~{\rm eV}$ cannot, at present, be used to infer robust constraints on ULDM.

\acknowledgments
We thank Asimina Arvanitaki, Joshua Eby, Reinhard Genzel, Rainer Sch\"odel and Wei Xue for useful discussions, and Vitor Cardoso and Sergey Sibiryakov for comments on the manuscript. We wish we could consult with Tal Alexander about S star orbits. KB is incumbent of the Dewey David Stone and Harry Levine career development chair. TL has received funding from the European Union's Horizon 2020 research and innovation programme under the Marie Sk\l{}odowska-Curie grant agreement No. 713366. The work of TL has also been supported by the Spanish Agencia Estatal de Investigaci\'{o}n through the grants PGC2018-095161-B-I00, IFT Centro de Excelencia Severo Ochoa SEV-2016-0597, and Red Consolider MultiDark FPA2017-90566-REDC. 

\begin{appendix}

\section{ULDM soliton near a black hole}\label{ss:solbh}

In this section we consider the ULDM soliton in the presence of a SMBH. In section \ref{sss:shape} we consider the basic features of the soliton solution, reviewing and extending an earlier analysis in Ref.~\cite{Bar:2018acw} (see appendix B there) and introducing some useful approximations. 
In section \ref{sss:absorb} we consider the accretion of mass from the soliton into the SMBH.

\subsection{Soliton shape}\label{sss:shape}

We consider a real, massive, free
scalar field $\phi$ satisfying the Klein-Gordon equation of motion (EoM)
and minimally coupled to gravity.\footnote{Analyses of interacting fields can be found in,
	e.g.~Refs.~\cite{Chavanis:2011zi,Chavanis:2011zm,RindlerDaller:2012vj,Desjacques:2017fmf}.}   
We look for a spherically-symmetric, quasi-stationary bound state solution in the non-relativistic regime. To this end we decompose $\phi$ as
\be\label{eq:schroedfield}\phi(x,t)=\frac{e^{-im(1+\gamma)t}}{\sqrt{8\pi G}}\chi(x)+\mathrm{c.c.} \ ,\ee
where $G$ is the gravitational Newton's constant and $\gamma$ is an eigenvalue of the problem. We do not need the explicit value of $\gamma$ in what follows. 

We rescale the spatial coordinate, 
\be  r&=& m\,x \ ,\ee
assuming spherical symmetry. 
The EoM for $\chi$ and for the Newtonian gravitational potential $\Phi$, sourced by $\chi$, is 
\be\label{eq:eom1}\partial^2_r\left(r\chi\right)&=&2r\left(\Phi-\frac{A}{r}-\gamma\right)\chi \ ,\\
\label{eq:eom2}\partial^2_r\left(r\Phi\right)&=&r\chi^2 \ ,\ee
where
\be\label{eq:A} 
\begin{split}
A=& \,\, G\MBH m \\
\approx& \,\, 3\times10^{-3}\left(\frac{\MBH}{4\times10^6~\Msol}\right)\left(\frac{m}{10^{-19}~\rm eV}\right) \ .
\end{split}
\ee

The lowest energy bound state solutions of Eqs.~(\ref{eq:eom1}-\ref{eq:eom2}) are parametrised by a single continuous positive parameter $\lambda$ that can be chosen as the value of the field at the origin: $\chi_\lambda(0;A)=\lambda^2$. Once we solve for $\chi_1$ and $\Phi_1$, all other solutions are obtained via
\be\chi_\lambda(r;A)&=&\lambda^2\chi_1(\lambda r;A/\lambda) \ ,\\
\Phi_\lambda(r;A)&=&\lambda^2\Phi_1(\lambda r;A/\lambda) \ .\ee
The mass density of the $\chi_\lambda$ soliton is
\be\rho_\lambda(r;A)&=&\frac{m^2}{4\pi G}\lambda^4\chi_1^2\left(\lambda r;A/\lambda\right).\ee
The total mass is given by
\be M_\lambda(A)&=&\lambda M_1(A/\lambda) \ ,\ee
where $M_1$ is the mass in $\chi_1$. In the limit $A\to0$, one finds 
\be M_1(0)&\approx&\frac{2.06}{Gm} \ .
\ee
In the opposite limit one finds 
\be M_1(A\gg1)&=&\frac{1}{4A^3Gm} \ .
\ee

\begin{figure}[htbp!]
	\centering
	\includegraphics[width=0.45\textwidth]{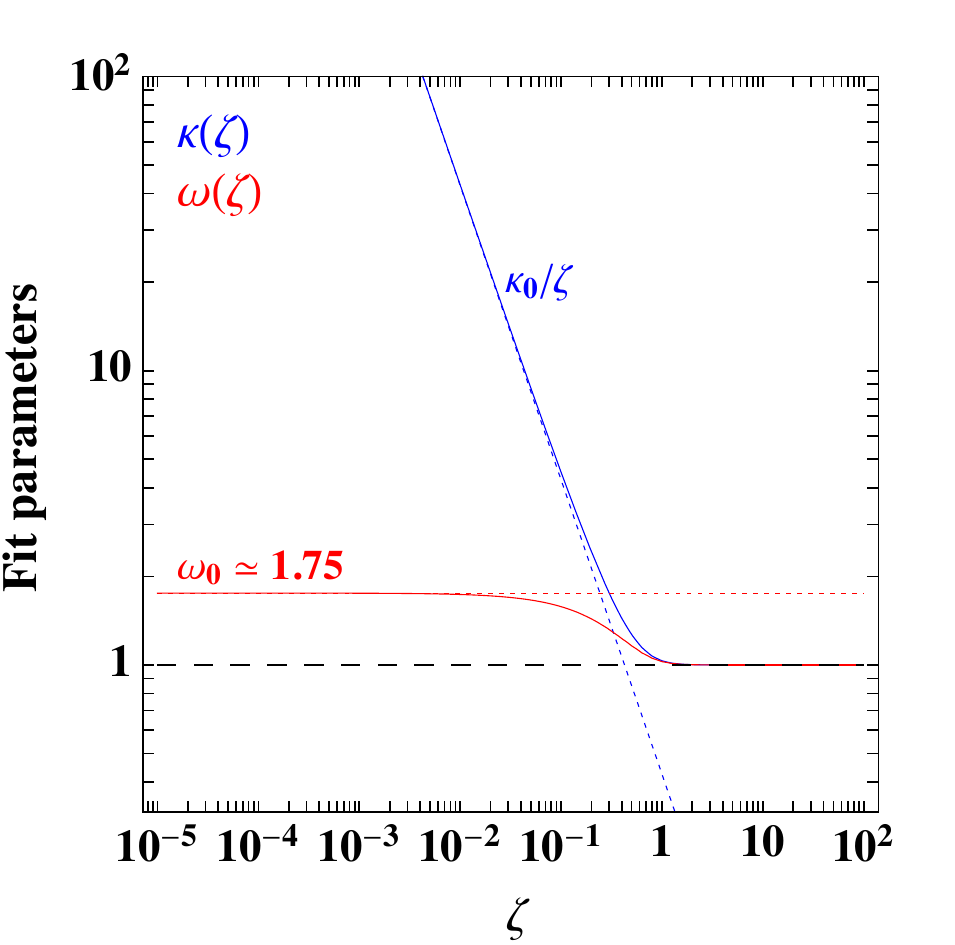}	
	\caption{Fit parameters $\kappa$ and $\omega$ for the soliton density profile.}\label{fig:FitParameters}
\end{figure}

Solving numerically for $\chi_1(r;A)$ is not particularly difficult. However, it is useful to find a semi-analytic approximation for the mass density $\rho_\lambda(r;A)$ that applies in both limits $A\to0$ and $A\gg1$. One such useful approximation is given by 
\be\label{eq:fit1}\rho_\lambda(r;A)&\approx&\frac{m^2}{4\pi G}\lambda^4\,e^{-2\left[\kappa(\zeta)\,A\,r\right]^{\omega(\zeta)}} \ ,\\
\label{eq:fit2}\zeta&\equiv&\frac{A}{\lambda} \ .\ee
The functions $\kappa(\xi)$ and $\omega(\xi)$ are found by a numerical fit to the exact solution, and are shown in figure \ref{fig:FitParameters}.
At $\zeta\gg1$ we have $\omega\to1$ and $\kappa\to1$, which reproduces the Coulomb solution, $\chi_1(r;A\to\infty)\to e^{-Ar}$, where the soliton profile is purely controlled by the BH gravity. At $\zeta\ll1$ we have $\omega\to \omega_0\simeq 1.75$, independent of $\zeta$, and $\kappa\to \kappa_0/\zeta \simeq 0.43/\zeta$. This gives a soliton profile that is independent of $A$ and reconstructs with good accuracy the self-gravitating exact numerical solution of $\chi_\lambda(r;0)$.
\begin{figure}[htbp!]
	\centering
	\includegraphics[width=0.45\textwidth]{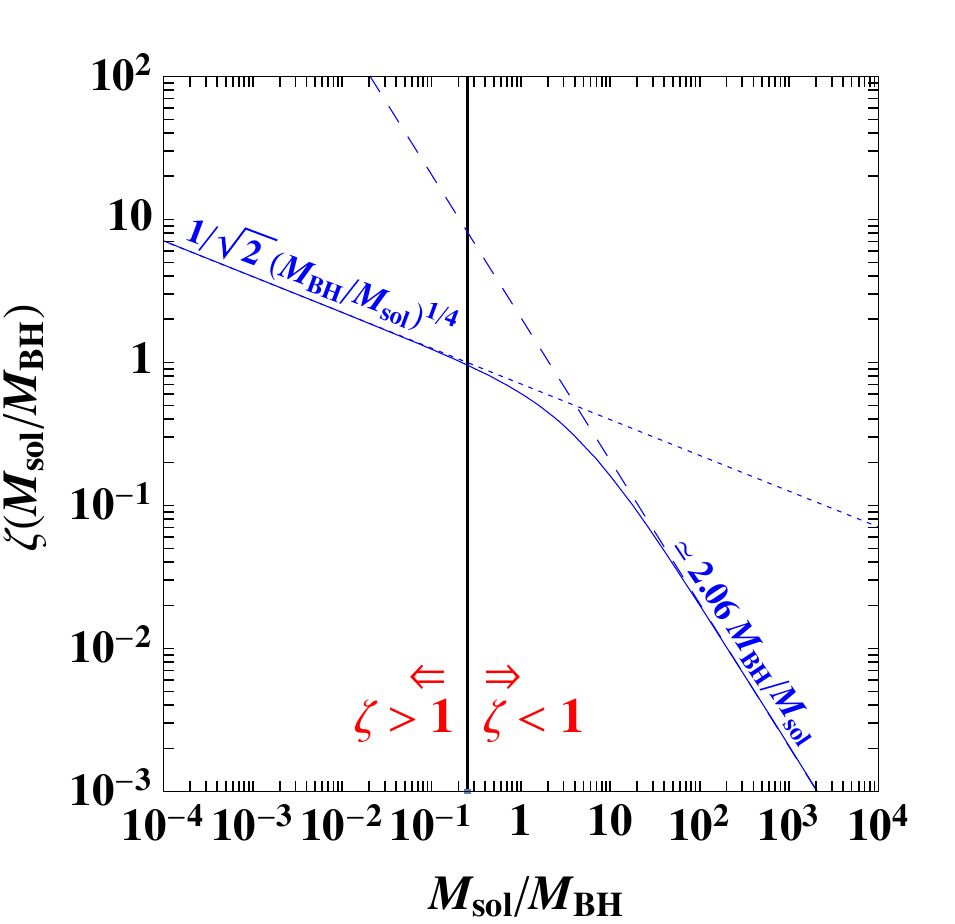}	
	\caption{Numerical fit of the parameter $\zeta$.}\label{fig:zeta}
\end{figure}

Using Eqs.~(\ref{eq:fit1}-\ref{eq:fit2}), the total mass of the $\lambda$-soliton reads
\be\label{eq:massfit} 
\begin{split}
M_\lambda(A) \approx & \, \frac{\lambda}{Gm}\left(\frac{1}{\zeta\,\kappa(\zeta)}\right)^3\frac{\Gamma\left(\frac{3}{\omega(\zeta)}\right)}{8^{\frac{1}{\omega(\zeta)}}\omega(\zeta)}\\
=& \, \lambda M_1(\zeta) \ .
\end{split}
\ee
From this expression we can obtain a numerical fit of the parameter $\zeta$, expressed as a function of the physical mass ratio $M_{\rm sol}/\MBH$. The results of this fit are shown in figure \ref{fig:zeta}. 
For later use we also compute the soliton mass $M^{\rm ext}_\lambda(A)$ enclosed in the shell between two radii $r_{\rm i}$ and $r_{\rm o}$ (corresponding to physical radii $x_{\mathrm{i},\mathrm{o}}=r_{\mathrm{i},\mathrm{o}}/m$):
\be M^{\rm ext}_\lambda(A)&\approx&M_\lambda(A)\left[\frac{\Gamma\left(\frac{3}{\omega(\zeta)},2\left[\kappa(\zeta)Ar\right]^{\omega(\zeta)}\right)}{\Gamma\left(\frac{3}{\omega(\zeta)}\right)}\right]\Big|_{r_{\rm i}}^{r_{\rm o}} \ .\no\\&&\ee

Finally, let us check the domain of validity of the Newtonian approximation of the EoM, which we have been using so far. The characteristic momentum associated with the ULDM field in the soliton is given by
\be\label{eq:k2}
\begin{split}
 k^2=& \, m^2\frac{\int dr\,r^2\,\left(\partial_r\chi\right)^2}{\int dr\,r^2\,\chi^2} \\ 
 \approx & \, \dfrac{\left[Am\,\kappa(\zeta)\omega(\zeta)\right]^22^{-2+\frac{2}{\omega(\zeta)}}\Gamma\left(2+\frac{1}{\omega(\zeta)}\right)}{\Gamma\left(\frac{3}{\omega(\zeta)}\right)} \ .
 \end{split}
 \ee
When the BH dominates the solution we find 
\be\label{eq:ktyp} k&\approx&A\,m \ .\ee
Similarly, the kinetic energy per unit mass is $ K/M\approx A^2/2 $ when the BH dominates the solution. Therefore, in the regime where the BH dominates the dynamics, the characteristic particle velocity associated with the soliton is given by $v\approx A$ and the Newtonian approximation remains adequate as long as $A\ll1$. Referring back to eq.~(\ref{eq:A}) we learn that for a SMBH with $\MBH=4\times10^6~\Msol$ (appropriate for Sgr A*) the Newtonian approximation breaks down for $m\gtrsim10^{-17}~{\rm eV}$, while for $\MBH=6.5\times10^9~\Msol$ (appropriate for M87*) the approximation breaks down already for $m\gtrsim10^{-20}~{\rm eV}$. Once we come close to the relativistic region, $A\sim1$, phenomenae such as superradiance come into play (see, e.g.~\cite{Arvanitaki:2010sy}). Since our analysis did not account for these effects, we stop our exploration of the parameter space at $m=10^{-18}~{\rm eV}$ and $m=10^{-21}~{\rm eV}$ in Figs.~\ref{fig:S2} and~\ref{fig:M87}, respectively.

\subsection{Absorption by the black hole}\label{sss:absorb}

A SMBH would accrete mass from the soliton and grow at its expense~\cite{Hui:2016ltb}. Ref.~\cite{Bar:2018acw} estimated the absorption rate of the soliton by the SMBH, in the limit that the soliton's self-gravity dominates over that due to the SMBH over most of the soliton region. Here we extend this estimate to the case where the SMBH gravity dominates throughout the soliton.

The cross section for
absorption of a scalar particle with mass $m$ and non-relativistic momentum $k\ll m$ by a
Schwarzschild BH, whose size is much smaller than the 
Compton wavelength of the particle, was calculated in Ref.~\cite{Unruh:1976fm}:
\be
\sigma_{\rm abs}&=&\frac{32\pi^2A^3}{k^2(1-e^{-\xi})} \ ,
\ee
where the $A$ parameter was defined in eq.~(\ref{eq:A}) and $\xi=2\pi Am/k$.
The mass accretion rate flowing from the soliton into the BH is then
\be\label{eq:Mdot}
\dot M& \approx &\frac{32\pi^2 A^3\rho_0}{mk(1-e^{-\xi})} \ ,
\ee
where $\rho_0$ is the soliton central density. Using our parametrisation, and combining Eqs.~(\ref{eq:fit1},\ref{eq:massfit},\ref{eq:k2}) with eq.~(\ref{eq:Mdot}), we can write the characteristic time for the SMBH to absorb the soliton
\be
t_{\rm abs} & \equiv & \dfrac{M_{\rm sol}}{\dot{M}} \\& =& \dfrac{\frac{\sqrt{\Gamma\left(2+\frac{1}{\omega}\right)\Gamma\left(\frac{3}{\omega}\right)}}{2^{4+\frac{2}{\omega}}\pi}\left[1-\exp\hspace{-1mm}\left({-\frac{\pi 2^{2-\frac{1}{\omega}}\sqrt{\frac{\Gamma\left(\frac{3}{\omega}\right)}{\Gamma\left(2+\frac{1}{\omega}\right)}}}{\kappa\omega}}\right)\right]}{\kappa^2 G^5 m^6 \MBH ^5} \ .\no
\ee

We are interested in the case where the BH dominates the soliton dynamics (the opposite limit was studied in Ref.~\cite{Bar:2018acw}), wherein $k=Am$ is given by eq.~(\ref{eq:ktyp}), leading to $\xi=2\pi$. Thus, we can neglect $e^{-\xi}$ in the denominator of eq.~(\ref{eq:Mdot}). In the same limit, the soliton central density is related to the total soliton mass $M_{\rm sol}$ by $\rho_0=(Am)^3M_{\rm sol}/\pi$. Combining these expressions with eq.~(\ref{eq:Mdot}), we find the characteristic time for the SMBH to absorb the soliton:
\be\label{eq:tabs} t_{\rm abs}&\equiv &\frac{M_{\rm sol}}{\dot M}\approx \frac{1}{32\pi A^5m}\\
&\approx&10\left(\frac{4\times10^6~\Msol}{\MBH}\right)^{5}\left(\frac{3\times10^{-20}~\rm eV}{m}\right)^{6}~{\rm Gyr}\no\\
&\approx&10\left(\frac{6.5\times10^9~\Msol}{\MBH}\right)^{5}\left(\frac{6.5\times10^{-23}~\rm eV}{m}\right)^{6}~{\rm Gyr}  .\no
\ee
This result is independent of the soliton mass.

In the second line of eq.~(\ref{eq:tabs}) the BH mass is scaled to that of Sgr A*, and in the third line it is scaled to that of M87*. In each line $m$ is scaled such that the soliton absorption time is about 10~Gyr. The strong dependence of $t_{\rm abs}$ on $m$ implies that for each of the two systems (Sgr A* and M87*), $m$ larger than the values shown in eq.~(\ref{eq:tabs}) would result in the soliton being rapidly consumed by the SMBH.

Throughout this section we have ignored corrections due to BH rotation. However, the results developed in~\cite{Benone2019} suggest that this may not lead to a large correction: for a Kerr BH with spin parameter $a$, using eq.~(51) in~\cite{Benone2019} we find
\be
\sigma_{\rm absorb} \to \sigma_{\rm absorb}\times \frac{1}{2}\left(1+\sqrt{1-a^2}\right) \ .
\ee
Thus, the absorption time scale $t_{\rm abs}$ could be longer by up to a factor of $2$ for a maximally rotating BH.

We conclude that much of the parameter space considered in the body of the paper may be strongly affected by BH absorption. This motivates a more careful analysis, that (i) goes beyond the single wavelength approximation of the soliton, adopted here for simplicity using the characteristic wavelength from eq.~(\ref{eq:k2}), and that (ii) solves the BH and the soliton co-evolution.

\section{Non-gravitational self-interactions}\label{app:selfint}
ULDM could be realised by axion-like particles (see, e.g.~\cite{Svrcek:2006yi}; for interesting alternatives, see~\cite{Davoudiasl:2017jke,Mishra2017}), with the relic abundance set by initial misalignment of the field in a cosine potential. In this case, non-gravitational self-interactions are expected to become significant in some regions of the $(m,M_{\rm sol})$ plane. 

Assuming the potential
\be
V(\phi)&=&m^2f^2\left[1-\cos\left(\frac{\phi}{f}\right)\right] \ ,
\ee
one finds the following self-interaction term $\delta V \approx -m^2\phi^4/24f^2  \equiv \kappa_{\rm I} \phi^4/4$ near the minimum of the potential, with $\kappa_{\rm I} = -m^2/6f^2$. 
In the misalignment mechanism, the ULDM relic abundance is given by $\Omega_{\rm dm}\sim 0.1\,(\phi_0/10^{17}~{\rm GeV})^2(m/10^{-22}~{\rm eV})^{1/2}$ where $\phi_0$ is the initial value of the field during inflation. Thus, the observed DM abundance is obtained for $\phi_0=10^{17} (m/10^{-22}~\rm{eV})^{-1/4}$~GeV. Let us define $\phi_0\equiv\theta f$, where a-priori one expects $\theta\sim\mathcal{O}(1)$, and fix $\phi_0$ as needed for the DM abundance. Plugging this into the expression for $\kappa_{\rm I}$ gives
\be
|\kappa_{\rm I}| \approx 1.7\times 10^{-97}\left(\dfrac{m}{10^{-22}~\text{eV}}\right)^{5/2}\theta^2 \ .
\ee

Self-interactions can be estimated to contribute a fraction $\sim\alpha$ of the energy density of the soliton solution when
\be
\dfrac{\left|\kappa_{\rm I}\right| \phi^4}{4} \sim \alpha \dfrac{|\nabla \phi|^2}{2} \ ,
\ee
where this estimate holds for $\alpha<1$. Noting that the ULDM mass density is given by $\rho = m^2\phi^2$, and utilising formulas from appendix \ref{sss:shape}, one finds that self-interactions contribute a fraction $>\alpha$ to the soliton energy density when
\be\label{eq:MsolSI}
M_{\rm sol} &\gtrsim& 2\pi\alpha \dfrac{M_{\rm pl}^2}{\MBH |\kappa_{\rm I}|\kappa}\dfrac{\Gamma\left(2+\frac{1}{\omega}\right)}{2^{\frac{1}{\omega}}} \ . 
\ee
In the BH-dominated case, this reduces to
\be
\begin{split}
M_{\rm sol} >&\, 2\pi\alpha \dfrac{M_{\rm pl}^2}{\MBH |\kappa_{\rm I}|} \\ 
 \approx & \, 2\times 10^{9}\dfrac{\alpha}{\theta^2}\left(\dfrac{10^{-21}~{\rm eV}}{m}\right)^{\frac{5}{2}}\left(\dfrac{6\times 10^{9}~M_{\odot}}{\MBH}\right) M_{\odot} \ .
\end{split}
\ee

In Figs.~\ref{fig:S2} and~\ref{fig:M87} we used a dashed magenta line to depict eq.~(\ref{eq:MsolSI}), setting $\alpha=0.1$ and $\theta=1$. From the equation above it is clear that smaller values of $\theta$ (for example $\theta=0.1$) corresponding to mild fine-tuning of the initial conditions, would push the onset of self-interactions two orders of magnitude up in $M_{\rm sol}$, making this effect irrelevant for the parameter space analyzed  in  this paper. Therefore, while it is interesting to contemplate what changes could occur in the soliton properties due to self-interactions via a generic cosine potential, we leave this investigation to other works.

\end{appendix}
\bibliography{ref}
\bibliographystyle{utphys}

\end{document}